\documentclass[aps,pra,pdf,superscriptaddress,twocolumn,showpacs,nofootinbib]{revtex4-2}
\usepackage{amsmath,amssymb,amsfonts}
\usepackage{eqnarray}
\usepackage{graphics,float}
\usepackage{bm}
\usepackage{braket}
\usepackage{amsmath}
\usepackage{physics}
\usepackage{mathtools}
\usepackage{gensymb}
\usepackage{color,xcolor,colortbl}
\usepackage{tikz}
\usepackage{soul}
\usetikzlibrary{shapes}
\usepackage[colorlinks=true,
linkcolor=blue,
filecolor=magenta,      
urlcolor=blue,
citecolor=blue]{hyperref}
\usepackage{lipsum}
\usepackage{orcidlink}
\usepackage[normalem]{ulem} 

\usepackage{cancel}

\definecolor{Gray}{gray}{0.85}
\definecolor{LightCyan}{rgb}{0.25,0.9,0.95}
\definecolor{LightMagenta}{rgb}{0.91,0.21,0.96}

\newcommand{\marker}[2]{%
\raisebox{-0.6pt}{%
\tikz{\node[draw,circle,line width=0.9pt,scale=0.6,#2,color=#1,fill=none]{};}%
}}

\definecolor{deepgreen}{RGB}{255,185,15}

\begin{document}
\title{Response of a dipolar BEC to Laguerre–Gaussian beam driven STIRAP}

\author{Deepu Singh\orcidlink{0009-0008-5809-0711}} 
\email{deepu.sy9880@gmail.com}
\affiliation{Department of Physics, Indian Institute of Technology Kharagpur, Kharagpur, West Bengal 721302, India}
\author{Hari Sadhan Ghosh\orcidlink{0009-0003-9969-390X}}

\affiliation{Department of Physics, Indian Institute of Technology Kharagpur, Kharagpur, West Bengal 721302, India}
\author{Arpana Saboo\orcidlink{0009-0000-8462-4547}}

\affiliation{Department of Physics, Indian Institute of Technology Kharagpur, Kharagpur, West Bengal 721302, India}
\author{Soumyadeep Halder \orcidlink{0000-0003-2824-8879}}

\affiliation{Department of Physics, Indian Institute of Technology Gandhinagar, Palaj, Gujarat 382055, India}

\author{Sonjoy Majumder\orcidlink{0000-0001-9131-4520}}
\email{sonjoym@phy.iitkgp.ac.in}
\affiliation{Department of Physics, Indian Institute of Technology Kharagpur, Kharagpur, West Bengal 721302, India} 

\begin{abstract}
    Coherent light–matter coupling via STIRAP can offer a versatile route to nucleate quantized vortices in Bose–Einstein condensates through the orbital angular momentum transfer from a vortex beam, yet its efficacy in dipolar condensates remains an open question. Can the orbital angular momentum of a Laguerre–Gaussian beam be coherently transferred to a dipolar BEC via STIRAP? We investigate this for a quasi-two-dimensional trapped dipolar condensate using co-propagating Gaussian and Laguerre–Gaussian laser beams. The interplay between long-range dipole–dipole interactions and short-range contact interactions enables access to three interaction-driven phases: superfluid, droplet, and supersolid. We find that the amount of angular momentum transferred from the optical field to the dipolar condensate, along with the nucleation and persistence of vortices, depends strongly on the underlying phases of the dipolar BEC. In the superfluid, STIRAP achieves a near-complete population transfer and nucleates a long-lived quantized vortex, reflecting efficient transfer of angular momentum to the condensate. In the droplet phase, although the vortex remains pinned within the density profile, the angular momentum is partially retained and oscillatory, accompanied by droplet fragmentation and recombination. In the supersolid phase, when the external magnetic field is oriented perpendicular to the LG beam's propagation direction, the emergence of a modulated density distribution along with a slight reduction in inter-droplet coherence leads to vortex delocalization and eventually exits from the condensate along the field direction, yielding a vanishing average angular momentum. However, reorienting the magnetic polarization along the beam propagation direction restores efficient angular momentum transfer and stabilizes the vortex within the supersolid phase. Our findings will pave the way for future experimental investigations of vortex nucleation in different phases of dipolar BECs via STIRAP-induced transfer of optical orbital angular momentum to the condensate and may serve as an important benchmark.
 
\end{abstract}

\maketitle

\section{Introduction}

Coherent manipulation of quantum states in ultracold atomic systems is an exciting area of research. The stimulated Raman adiabatic passage (STIRAP) technique \cite{vitanov_stimulated_2017, shore_picturing_2017} has emerged as an efficient tool for coherent-state engineering, facilitating population transfer between two quantum states with near-unity efficiency while avoiding occupation of an intermediate excited state \cite{zeng_nonclassical_1995, shore_picturing_2017, nandi_vortex_2004}. In addition to enabling population transfer between quantum states, STIRAP can coherently transfer angular momentum from laser fields to atoms, molecules, and Bose–Einstein condensates (BECs) \cite{Babiker_2019, andersen_quantized_2006, Pradip_2014, chen_spin--orbital-angular-momentum_2018, zeng_nonclassical_1995, nandi_vortex_2004, mukherjee_dynamics_2021, alexandrescu_electronic_2005, mukherjee_interaction_2018, Anal_magicwavelength_2018}. This transfer of angular momentum relies on the fact that light can carry two different types of angular momentum: spin angular momentum, associated with its polarization \cite{mkhumbuza_topological_2026}, and orbital angular momentum (OAM), associated with its spatial mode structure \cite{mkhumbuza_topological_2026}. For instance, a light beam with a phase singularity, such as a Laguerre-Gaussian (LG) beam, carries a well-defined orbital angular momentum $l\hbar$ per photon, where $l$ is the integer winding number of the phase of light \cite{mkhumbuza_topological_2026, Babiker_2019}. Such an LG beam can be used as either the pump or the Stokes beam in a STIRAP scheme. During the two-photon Raman transition, the system absorbs a photon from the pump laser and emits a photon into the other. The difference in angular momentum between the two photons is then transferred to the atom or molecule \cite{Babiker_2019, Pradip_2014, andersen_quantized_2006, araoka_interactions_2005}. In a BEC, STIRAP can facilitate the collective transfer of the OAM from a LG beam to the condensate. Owing to the single-valued nature of the condensate wave function and the spontaneously broken $U(1)$ gauge symmetry, the transferred angular momentum leads to the nucleation of quantized vortices in the condensate \cite{chen_spin--orbital-angular-momentum_2018, mukherjee_dynamics_2021, Pradip_CD_2015, Anal_NP_2016, bhowmik_role_2022, Das_2020, Bhowmik_2018, Bhowmik_2018_2222, Mukherjee_2018, shin_dynamical_2004}.\par 

In recent years, a new class of ultracold quantum gases composed of Dy or Er atoms with strong magnetic dipole moments has gained significant attention \cite{halder_control_2022, Casotti2024, klaus_observation_2022}. In addition to the standard isotropic contact-type interaction in typical BEC \cite{liu_fluctuation-driven_2024, mohamed_generalized_2023, adusumalli_quantum_2024, zhang_density-functional_2023, liu_fluctuation-driven_2024, petrov_quantum_2015, liu_one-dimensional_2022, Boudjemaa2021, liu_fluctuation-driven_2024, semeghini_self-bound_2018, PhysRevA.106.023306}, these dipolar condensates also enable long-range, anisotropic dipole-dipole interactions (DDI) \cite{halder_control_2022, halder_induced_2024, halder_two-dimensional_2023, bottcher_transient_2019, ferrier-barbut_observation_2016, chomaz_long-lived_2019, tanzi_supersolid_2019} between the atoms. The interplay between these two interactions can induce spontaneous breaking of continuous translational symmetry via the roton softening and the emergence of supersolid and droplet phases \cite{PhysRevLett.90.250403, lima_beyond_2012, pastukhov_beyond_2017, boudjemaa_theory_2014, pedri_two-dimensional_2005, ticknor_anisotropic_2011, fischer_stability_2006}, with the resulting state being stabilized by the effect of quantum fluctuations \cite{lima_beyond_2012, PhysRevA.84.041604, pastukhov_beyond_2017, boudjemaa_theory_2014, wenzel_striped_2017, Fischer1}. Both theoretical and experimental studies have demonstrated that vortices can be nucleated in the various phases of a dipolar condensate \cite{Casotti2024, klaus_observation_2022, poli_synchronization_2025}, including the superfluid, supersolid, and droplet phases via phase imprinting \cite{8n5y-fyh7, li_strongly_2024, PhysRevA.98.023618}, trap rotation \cite{PhysRevA.83.033628, PhysRevA.111.023301, halder_control_2022, PhysRevA.110.033322, PhysRevLett.88.010405, Sabari_2024, Mukherjee2023}, or magneto-stirring techniques \cite{Casotti2024, klaus_observation_2022, poli_synchronization_2025, halder_control_2022, Mukherjee2023}. The STIRAP protocol also offers a promising route for coherently transferring angular momentum from light to the various phases of dipolar condensates, thereby enabling the nucleation of quantized vortices. Although this technique has been extensively studied and implemented in non-dipolar BECs \cite{zeng_nonclassical_1995, shore_picturing_2017, nandi_vortex_2004, Babiker_2019, andersen_quantized_2006, Pradip_2014, chen_spin--orbital-angular-momentum_2018, nandi_vortex_2004, mukherjee_dynamics_2021, cappellaro_equation_2017, marzlin_vortex_1997, nandi_vortex_2004}, its impact on the different phases of dipolar condensates remains largely unexplored.\par

Recent experimental progress in the creation and stabilization of long-lived spin mixtures of ultracold dipolar Bose gases \cite{lecomte_production_2025}, along with theoretical predictions of various unique mixed-phases in binary dipolar BEC \cite{halder_induced_2024, halder_two-dimensional_2023, bisset_quantum_2021, smith_quantum_2021, zhang_metastable_2024, bland_alternating-domain_2022, scheiermann_catalyzation_2023}, has been enabled by the suppression of dipolar relaxation and controlled tuning of short-range interactions. These advances open the possibility of implementing STIRAP-based coherent-state engineering protocol in dipolar condensates, thereby providing a unique platform for investigating optical OAM transfer under the influence of anisotropic long-range intra- and interspecies DDI, and pave a new route to form vortices in different phases of dipolar BECs \cite{sabari_vortex_2024, klaus_observation_2022}.\par

\begin{figure}[t] 
    \centering
    \includegraphics[width= \linewidth, height=4.2cm]{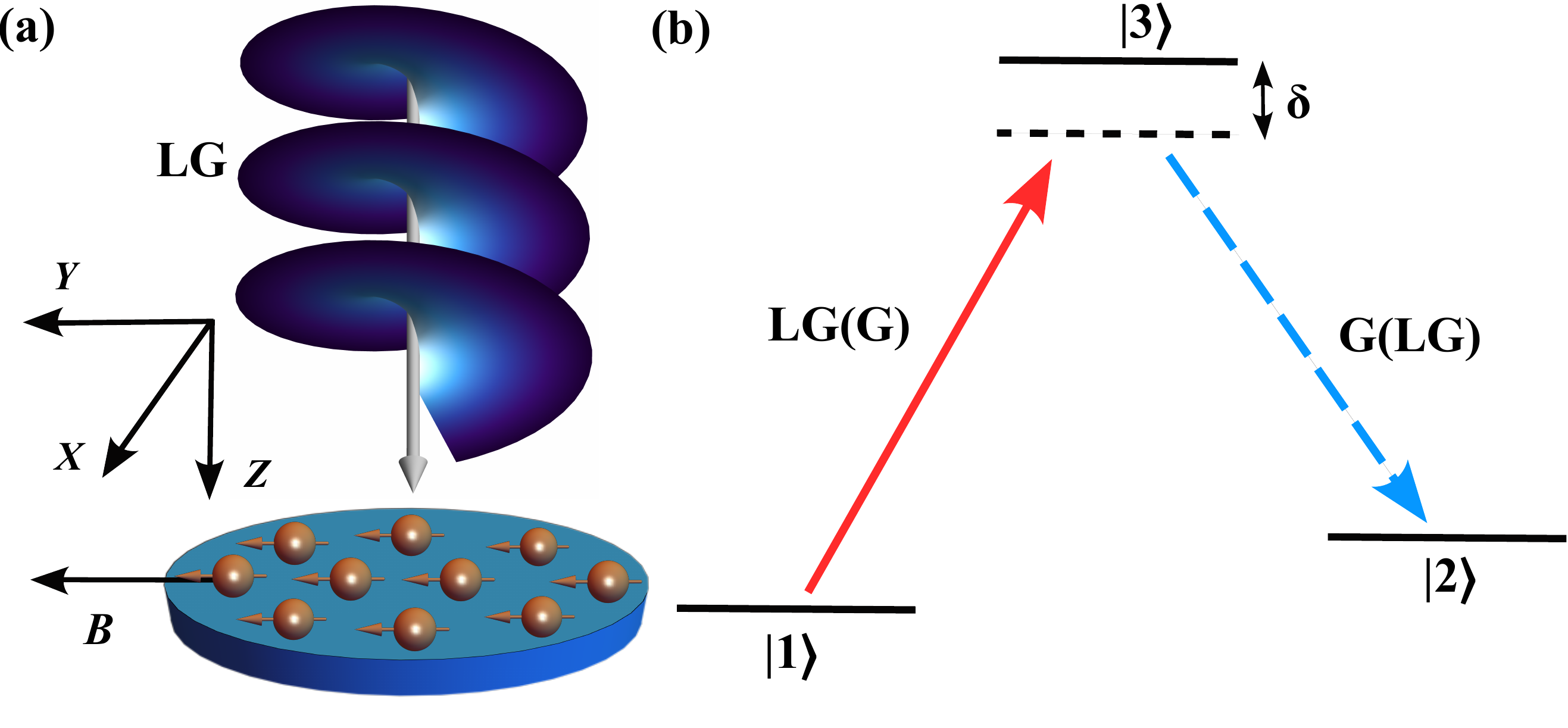}
    \caption{(a) Schematic illustration of LG beam propagating along the $z$-axis and interacting with a dipolar BEC confined in the $x$-$y$ plane. The dipoles are aligned by an external magnetic field applied along the $y$-direction. (b) Schematic illustration of the $\Lambda$-type STIRAP protocol. The red solid arrow corresponds to the Pump laser [LG(G)] driving the $\ket{1}\rightarrow\ket{3}$ transition, whereas the blue dashed arrow represents the Stokes beam [G(LG)] driving the $\ket{3}\rightarrow\ket{2}$ transition. The excited state $\ket{3}$ is detuned by $\delta$ from resonance. The hyperfine levels $\ket{8,-8}$ and $\ket{8,-7}$ of $^{162}$Dy are simply denoted as $\ket{1}$ and $\ket{2}$ \cite{lecomte_production_2025}, respectively. The counterintuitive sequence of Pump and Stokes pulses enables coherent population transfer between the two hyperfine states $\ket{1}$ and $\ket{2}$ while keeping the state $\ket{3}$ unoccupied \cite{mukherjee_dynamics_2021}.}
    \label{fig:1} 
\end{figure}

Motivated by these recent developments, in this article, we investigate the feasibility of transferring the OAM of light to different DDI-driven phases of dipolar BECs and the associated particle transfer dynamics between hyperfine states using the STIRAP protocol. To implement the STIRAP scheme, we employ a pair of copropagating Laguerre–Gaussian (LG) and Gaussian (G) beams as the Pump and Stokes beams, and vice versa, as illustrated schematically in Fig.~\ref{fig:1}. Furthermore, to stabilize the DDI-dominated phases against mean-field-driven collapse, we incorporate the effect of quantum fluctuations via the quasi-two-dimensional Lee-Huang-Yang (LHY) correction for a binary dipolar mixture, enabling us to access droplet and supersolid configurations within our theoretical framework. \par

The remainder of this paper is organized as follows. In Sec.~\ref {sec:theory}, we outline the theoretical model for the OAM transfer mechanism in a dipolar BEC. In Sec.~\ref{sec:numerical} describes the numerical methods used. In Sec.~\ref{sec:results}, we show that the interaction-driven phases (superfluid, supersolid, and droplet phase) of the dipolar Bose gas influence how the OAM is transferred, and whether efficient OAM is possible across these distinct phases. We further present our results on population transfer, showing that the Raman coupling and the particle interaction strength determine complete population transfer. In Sec.~\ref{sec:conclusion}, we summarize our findings and draw important conclusions. Appendix~\ref{app:eom} presents the derivation of the underlying equations of motion from the system Hamiltonian, Appendix~\ref{app:lhy} details the derivation of the quasi-2D LHY correction, and Appendix~\ref{app:modes} discusses collective excitations such as scissors and quadrupole modes.

\section{Theory}
\label{sec:theory}
We study the STIRAP-mediated optical OAM transfer in different phases of a quasi-2D dipolar BEC by using co-propagating Laguerre–Gaussian (LG) and Gaussian (G) beams as either the Pump or the Stokes beam. The system is subject to strong confinement along the $z$-direction, as shown schematically in Fig.~\ref{fig:1}(a). The condensate is initially prepared in the hyperfine state $\ket{1}$. The state $\ket{3}$ acts as an excited intermediate state and is far off resonance with a large single photon detuning $\delta$. The state $\ket{2}$ corresponds to the final hyperfine state populated through the STIRAP process [see Fig. \ref{fig:1}(b)]. The choice of LG and G beam as either the Pump or Stokes beam gives rise to two distinct physical situations. In general, if the Pump beam carrying OAM of $l_1\hbar$ drives the $\ket{1}\to\ket{3}$ transition and the Stokes beam carrying OAM of $l_2\hbar$ couples the $\ket{3}\to\ket{2}$ transition, the atoms transferred to the final state $\ket{2}$ via the STIRAP acquire a net angular momentum of $(l_1-l_2)\hbar$. Although the dipole coupling between the $\ket{1}$ and $\ket{2}$ states is mediated through the intermediate state $\ket{3}$, the large detuning $\delta$ ensures that $\ket{3}$ remains essentially unoccupied throughout evolution. This allows for the adiabatic elimination of the intermediate state $\ket{3}$. Under these conditions, coherent population transfer from $\ket{1}$ to $\ket{2}$ occurs.

The electric fields  of the pump and Stokes beams can be expressed as 

\begin{align}
    {\bf E}_j(\mathbf{r} , t) =& \mathbf{\hat{e}}_j \mathcal{E}_j(t) r_{\perp}^{\lvert l_j \rvert} \exp{-r_{\perp}^2/{w_j^2}}\nonumber \\ &\times\exp{-\mathrm{i}(k_{j}z-l_j\varphi - \omega_{j} t)},
    \label{electricfiled}
\end{align}
where $\mathbf{\hat{e}}_j$, $\mathcal{E}_j(t)$, $k_j$, and $\omega_j$ denote the unit polarization vector, time-dependent amplitude, wave number, and frequency of the $j$-th pulse, respectively,  with $j =1, 2$ denoting the pump and Stokes fields. The radial coordinate in the transverse plane is given by $r_\perp = \sqrt{x^2 + y^2}$, $w_j$ represents the beam waist, and $\varphi$ denotes the azimuthal angle. 

The temporal envelopes of the pulses are taken to be Gaussian in the form
\begin{align}
    \mathcal{E}_j(t) = E_{\rm{max}} \exp{-\frac{(t - \tau_{j})^2}{T^2}},
    \label{temporal}
\end{align}
where $E_{\text{max}}$ is the peak amplitude, $\tau_{j}$ is the temporal position of the pulse-centre , and $T$ is the pulse duration \cite{mukherjee_dynamics_2021, ghosh_dastidar_pattern_2022}. \par

During coherent population transfer, the condensate will become a mixture of the same species but different hyperfine levels (two-component BEC).
The coherent dynamics of the condensate components in the hyperfine states $\ket{1}$ and $\ket{2}$ under the STIRAP protocol are described by the macroscopic wave functions $\Psi_1(\vb{r}_{\perp}, t)$ and $\Psi_2(\vb{r}_{\perp}, t)$, respectively. Their temporal evolution follows a set of Raman-coupled extended Gross–Pitaevskii equations (eGPE) of the following form

\begin{align}
&\mathrm{i} \hbar \partial_t \Psi_i(\vb{r}_{\perp}, t) = 
\Biggl[
- \frac{\hbar^{2}}{2m_i} \nabla_{\perp}^2 + V^{\rm{eff}}_{i}(\vb{r}_{\perp}) + 
(\Delta \mu_i )^{\mathrm{LHY}}_{\rm 2D} \nonumber \\
&+ \sum_{j=1}^2 \biggl( G_{ij} \abs{\Psi_{j}(\vb{r}_{\perp}, t)}^2  
 + c^{ij}_{\rm{dd}} \mathcal{F}^{-1}_{2\rm D} \{ \tilde{n}_{j}(\vb{k}_{\perp},t) F(\vb{k_\perp} l_z / \sqrt{2}) \}\biggr)
\Biggr] \nonumber \\
&\times\Psi_i(\vb{r}_{\perp}, t) + 
\mathcal{V}'(\vb{r_{\perp}},t)\Psi_{3-i}(\vb{r_{\perp}}, t)e^{(-1)^{i}\mathrm{i}(l_1-l_2)\varphi},
\label{eq:1}
\end{align} 
where $i=1,2$. Here, light fields generate an additional spatial potential, leading to  an effective potential for atoms in the $i$-th component as

\begin{equation}
    V^{\rm{eff}}_{i} (\vb{r}_{\perp}) = V_i(\vb{r}_{\perp}) + \mathcal{V}_{i}(t)r_{\perp}^{2|l_{i}|} e^{- 2r_{\perp}^2 / w_{i}^2},
    \label{eq:2}
\end{equation}
where $V_i(\vb{r}_{\perp}) = \frac{1}{2}m_i \omega^{2}(x^2 + \kappa^{2} y^{2})$ is the asymmetric 2D  harmonic trapping potential with $\kappa$ as the trap anisotropy parameter. Here, since we consider the same species in different hyperfine levels, the mass of the atom $m_1=m_2=m$.
The corresponding harmonic oscillator lengths are $l = \sqrt{\hbar / m\omega}$ and $l_z = \sqrt{\hbar / m\omega_z}$ for the radial and axial directions, respectively.
The spatially dependent Raman coupling strength is given by 
$\mathcal{V}'(\mathbf{r_{\perp}}, t)= \sqrt{\mathcal{V}_{1}(t) \mathcal{V}_{2}(t)} (r_{\perp})^{(|l_1| + |l_2|)} 
e^{-(r_{\perp}^2){\left( 1 / w^2_1 + 1 / w^2_2 \right)}}$, where  $\mathcal{V}_i(t) = \mathcal{V}_{\rm{max}}  e^{-2\left[ (t - \tau_i) / T \right]^2}$ and $\mathcal{V}_{\rm{max}} =  E^{2}_{\rm{max}} d^{2} / \hbar\delta$ and $d$ is the induced electric dipole moment.

The coupling constant $G_{ij} = 2\sqrt{2\pi \lambda} a_{ij} \hbar^{2}/ m l$ in Eq.~\eqref{eq:1} represents the short-range intra- ($i=j$) and inter-species ($i\ne j$) contact interaction strengths, where $\lambda = \omega_{z}/ \omega$ is the trap aspect ratio with $\omega_{z}$ as trap frequency along $z$-direction and $a_{ij}$ is the $s$-wave scattering length. The dipole-dipole interaction is calculated first in the Fourier domain and transferred back to the real space using the inverse Fourier transformation in the quasi-2D frame, and obtained $\mathcal{F}^{-1}_{2\mathrm{D}}$. The function $F(\mathbf{q})$ represents the DDI in momentum space for a quasi-2D geometry \cite{nath_phonon_2009, raghunandan_two-dimensional_2015, shen_properties_2018, pedri_two-dimensional_2005, ticknor_anisotropic_2011, klawunn_hybrid_2009, fischer_stability_2006} with its coupling strength $c^{ij}_{\mathrm{dd}} = \mu_0 \mu_{i}^m \mu_j^m / (3\sqrt{2\pi} l_z)$, where $\mu_0$ is the vacuum permeability and $\mu_i^m ~(i=1,2)$ is the magnetic moment, and $\mathbf{q} = \mathbf{k_\perp} l_z / \sqrt{2}$. $F(\mathbf{q}) = \cos^2(\alpha) F_\perp(\mathbf{q}) + \sin^2(\alpha) F_\parallel(\mathbf{q}),$ where $\alpha$ is the angle between the polarization unit vector $\hat{\mathbf{d}}$ and the $z$-axis, $F_\perp(\mathbf{q}) = 2 - 3\sqrt{\pi} \, q e^{q^2} \operatorname{erfc}(q),$  and $F_\parallel(\mathbf{q}) = -1 + 3\sqrt{\pi} \frac{q_d^2}{q} e^{q^2} \operatorname{erfc}(q),$ where $\mathbf{q}_d$ is the wave vector along the direction of the projection of $\hat{\mathbf{d}}$ onto the $x$-$y$ plane, and $\operatorname{erfc}$ representing the complementary error function. Here, $\tilde{n}_{j}(\mathbf{k_\perp},t)$ is the 2D Fourier transformation of density $n_{j}(\mathbf{r_\perp},t)$.

In this work, we investigate the impact of the STIRAP protocol on dipolar BEC phases, including the dominant dipole-dipole interaction-driven density modulated phases, such as the droplet and supersolid phases. The mean-field theory fails to explain the stability of these phases against collapse, compelling the LHY correction crucial for the realization of stable droplets and supersolids. The third term in Eq.~\eqref{eq:1} accounts for the correction in the chemical potential due to the quantum fluctuations (known as LHY correction), derived in Appendix~\ref{app:lhy} for the quasi-2D case, is expressed as
\begin{equation}
(\Delta \mu_i )^{\mathrm{LHY}}_{2\mathrm{D}} = \frac{m^{3/2} \lambda^{3/4}}{3\sqrt{5}\pi^2 \hbar^3 l^{3/2} \pi^{3/4}} \sum_{\pm} \int_0^{1} \dd{u}  Re(I^\prime_{E \pm}),
\label{eq:3}
\end{equation}
where 
\begin{align}
&I^\prime_{E \pm} =  \left( \Tilde{V}_{ii}^{int}(u) \pm \frac{(-1)^{i-1} \delta \Tilde{V}_{ii}^{int}(u) + 2\Tilde{V}_{12}^{int}(u)^2 n_{3-i}}{\sqrt{\delta^2 + 4\Tilde{V}_{12}^{int}(u)^2 n_1 n_2}} \right) \nonumber \\
& \times\bigg(n_1 \Tilde{V}_{11}^{int}(u) +  n_2 \Tilde{V}_{22}^{int}(u) \pm \sqrt{\delta^2 + 4 \Tilde{V}_{12}^{int}(u)^2 n_1 n_2} \bigg)^{3/2}.
\label{eq:4}
\end{align}
$\delta = n_1 \Tilde{V}_{11}^{int}(u) - n_2 \Tilde{V}_{22}^{int}(u), \, \Tilde{V}_{ij}^{int}(u)=\Tilde{g}^{c}_{ij} + \Tilde{V}_{\rm{dd}}^{ij}(u),$ being the 2D Fourier transform of the total interaction potential.

Initially, in the absence of the optical fields of lasers, all atoms are assumed to occupy the lowest hyperfine state $\ket{1}$ and form a condensate. Upon interaction with the optical fields, the atoms are coherently transferred from state $\ket{1}$ to $\ket{2}$ via a STIRAP as described in Eq.~\eqref{eq:1}. At any instant of time, during the population transfer total number of atoms are conserved, i.e., $\int \big(|\Psi_1|^2 + |\Psi_2|^2\big)d\mathbf{r}_{\perp} = N_1+N_2=N$,  where $N_{1}$ and $N_{2}$ denote the number of atoms in states $\ket{1}$ and $\ket{2}$, respectively.

The transferred atoms acquire an orbital angular momentum of $(l_{1}-l_{2})\hbar$, represented as a phase factor $e^{\mathrm{i}(l_{1}-l_{2})\varphi}$ in the wavefunction $\Psi_{2}$. This transfer of OAM from light to the condensate $\Psi_{2}$ results in the nucleation of quantized vortices. Due to the single-valuedness of the condensate wavefunction, the phase must satisfy the quantization condition of an integer multiple of $2\pi$, characterized by a phase winding of $\mathcal{L}_q\times2\pi$, where $\mathcal{L}_q = l_{1}-l_{2}$ denotes the vortex topological charge. 

\section{Numerical Methods}
\label{sec:numerical}

In this work, following the STIRAP protocol, the dynamics are governed by numerical simulations of the set of Raman-coupled eGPE~\eqref{eq:1}. For the sake of convenience of numerical simulations and better computational precision, we cast Eq.~\eqref{eq:1} into a dimensionless form. This is achieved by rescaling the length scale and time scale in terms of oscillator length $l$ and trapping frequency $\omega$. Under this transformation, the wave function of the condensate in the hyperfine state $\ket{i}$ obeys $\Psi_i^{\prime}(\vb{r}_\perp)=\sqrt{l^2/N}\Psi_i(\vb{r}_\perp)$, where $N$ is the total number of particles in the state $\ket{1}$ and $\ket{2}$. After the transformation of variables into dimensionless quantities, the coupled eGPE~\eqref{eq:1} is solved by the split-step Crank-Nicolson scheme \cite{crank_1947_practical}. \par 
First, the initial configuration of the ground state is obtained by propagating Eq.~\eqref{eq:1} in imaginary time until the relative deviation of the condensate wave function, $\Psi_1$, between successive time steps becomes smaller than $10^{-6}$, while $\Psi_2$ is considered zero. This solution is then used as the initial state for the dynamical simulations, where Eq.~\eqref{eq:1} is propagated in real time following the STIRAP protocol. The STIRAP process is implemented using a combination of G-LG or LG-G pulse sequences acting as the pump and Stokes beams. The widths of both pulses are fixed at $T = 4\omega^{-1}$, with $\tau_{1} = 9.9\omega^{-1}$ and $\tau_{2} = 1.4\omega^{-1}$. The LG beam has a waist $w = 15l$. Our simulations are performed within a 2D square grid containing $(512\times 512)$ grid points with the spatial grid spacing $\Delta x=\Delta y=0.08 l$ while the time step $\Delta t= 2 \times 10^{-4}/\omega$.

\section{Results and Discussion}
\label{sec:results}
In this work, we consider a dipolar BEC of $^{162}$Dy atoms with the lowest two hyperfine states $|8,-8\rangle$ and $|8,-7\rangle$ defined as the initial $\ket{1}$ and final $\ket{2}$ states \cite{lecomte_production_2025}, respectively. The dipole moments corresponding to these hyperfine states are $-9.93\,\mu_B$ and $-8.69\,\mu_B$, respectively. Atoms with the $\ket{8,-8} \equiv \ket1$ state are prepared in the presence of an external magnetic field along the $y$-axis and subsequently investigated their response under the STIRAP protocol governed by Eq \eqref{eq:1} with the sequence of LG and G pulses, as well as the corresponding G-LG configuration. The system consists of $N = 6 \times 10^{4}$ number of dipolar atoms confined in a circularly symmetric quasi-2D harmonic trap with a relatively smaller trapping frequency in the $x$-$y$ plane $\omega = 2\pi \times 45$ Hz and the system experiences strong confinement along the $z$-direction, characterized by $\lambda = \omega_z/\omega = 20$ \cite{raghunandan_two-dimensional_2015, shen_properties_2018, pedri_two-dimensional_2005, ticknor_anisotropic_2011, klawunn_hybrid_2009, fischer_stability_2006}. The quasi-2D condition, $\hbar \omega_{z} \gg \mu_{1(2)}$, is maintained throughout the process. The transfer of angular momentum and the dynamical evolution of the condensate are sensitive to both intra- and inter-species DDI  as well as the contact-interaction strengths. A harmonically trapped dipolar BEC can exhibit superfluid (SF), supersolid (SS) and isolated droplet (ID) phases \cite{halder_control_2022, halder_induced_2024, halder_two-dimensional_2023} depending on the relative strength of DDI over the contact interaction strength, which can be achieved experimentally \cite{bottcher_transient_2019, ferrier-barbut_observation_2016, chomaz_long-lived_2019, tanzi_supersolid_2019, wenzel_striped_2017}.

\subsection{Superfluid Phase}
\label{subsec:superfluid}
\begin{figure*}[t]
    \includegraphics[width=\textwidth]{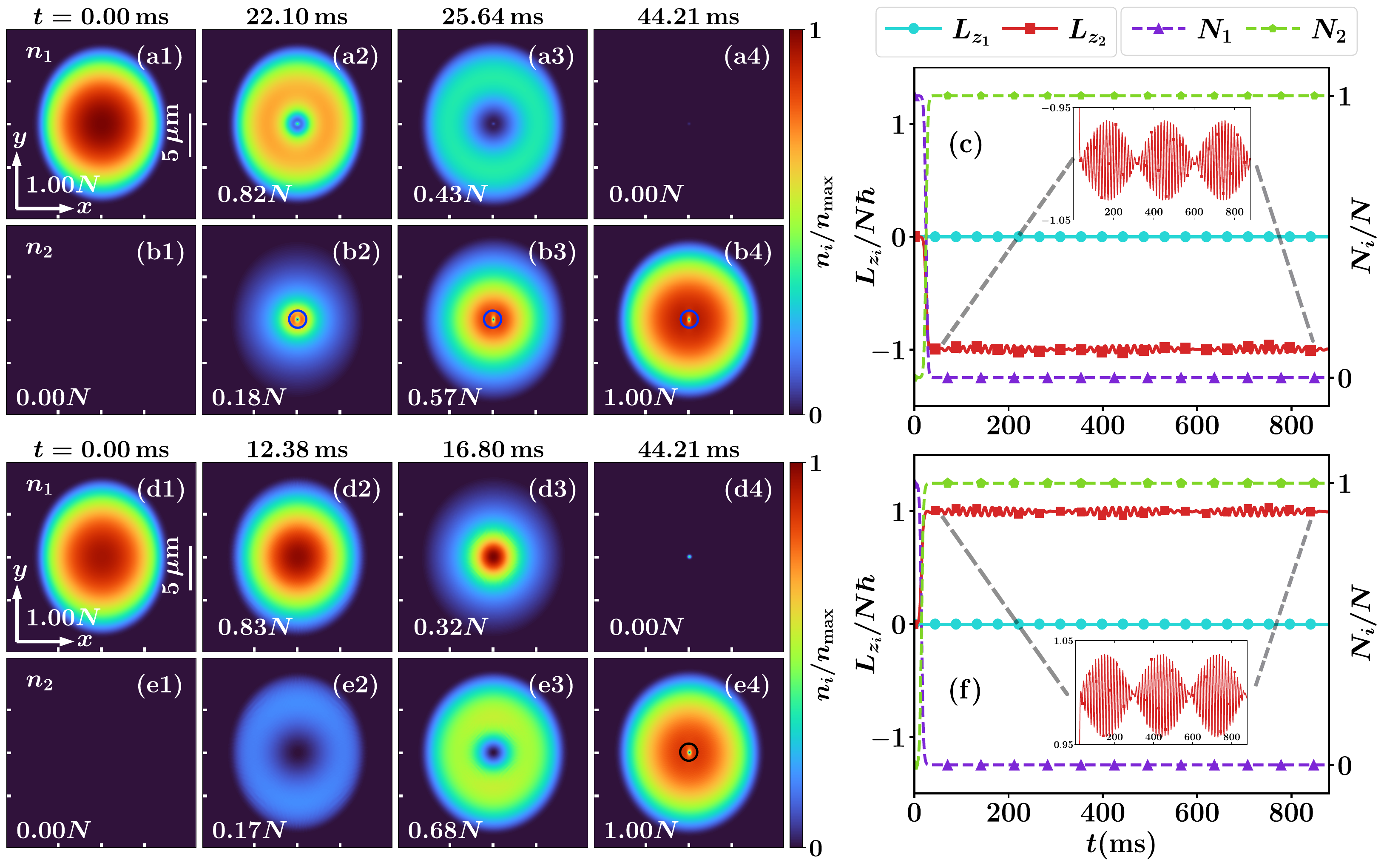}
    \caption{Snapshots of the condensate density distributions during the STIRAP-induced population transfer dynamics in the SF phase with dipole polarization along the $y$-axis. Panels (a1–a4) and (b1-b4) represent the density profiles of states $\ket{1}$ and $\ket{2}$, respectively, for the G-LG pulse sequence, while panels (d1–d4) and (e1-e4) show the density profiles of states $\ket{1}$ and $\ket{2}$, respectively, for the LG-G pulse sequence. The blue circle \protect\marker{blue}{} and the black circle \protect\marker{black}{} mark the position of the anti-vortex and vortex, respectively. Variation of the angular momentum $L_{z_i}/N\hbar$ and fraction of atom population $N_i/N$ in both states is shown in panel (c) for the G-LG pulse sequence and in panel (f) for the LG-G pulse sequence. The insets in (c) and (f) display an enlarged view of $L_{z_i}/N\hbar$. 
    }
    \label{fig:2}
\end{figure*}
We first examine the impact of the STIRAP protocol on the SF phase. To realize the SF phase, we consider relatively large intraspecies scattering lengths for atoms in both the hyperfine levels $\ket{1}$  and $\ket{2}$. In particular, we set $a_{11}=a_{22}=140a_0$, which are larger than the corresponding intraspecies dipolar lengths, $a_{11}^{\rm{dd}}=129.28a_0$, $a_{22}^{\rm{dd}}=99a_0$. Furthermore, the interspecies $s$-wave scattering lengths are also considered such that $a_{12}=a_{21}=150a_0$ exceeds the corresponding interspecies dipolar length $a_{12}^{\rm dd}=113a_0$. The coherent population transfer from $\ket{1}$ to $\ket{2}$ controlled by STIRAP pulse sequence on the initially prepared SF state, shown in Fig.~\ref{fig:2}(a1), with a Gaussian beam ($l_1{=0}$) and a Laguerre-Gaussian beam ($l_2{=}1$) as pump and Stokes beams, respectively, with peak amplitude $\nu_{\max}=100\,\hbar\omega$.
The Gaussian pump beam generates an effective repulsive potential around the center of $\Psi_1$ [see Eq.~\eqref{eq:2}]. It initiates a coherent transfer of atoms via a STIRAP primarily from the central region of $\Psi_1$. As the system evolves over time, atoms are progressively depleted from state $\ket{1}$ and populate state $\ket{2}$. During population transfer, the system exhibits a mixture of transient immiscible superfluid phase, as seen in the density profiles in Figs.~\ref{fig:2}(a2-a3) and \ref{fig:2}(b2-b3). The spatial dependence of the light-induced Raman coupling term in Eq.~\eqref{eq:1} enforces density depletion at the center $(r_{\perp}=0)$ of $\Psi_2$, while the associated phase factor transfers angular momentum, leading to the formation of an anti-vortex in the second component. The associated evolution of the angular momentum and the number of particles in each hyperfine state is shown in Fig.~\ref{fig:2}(c). Fluctuations in $L_{z_2}/N$ about $-\hbar$ are attributed to the excitation of the scissors and quadrupole modes during the population and angular momentum transfer dynamics (see Appendix~\ref{app:modes}). The nearly complete population transfer from $\ket{1}$ to $\ket{2}$ is achieved by $t=44.21~\mathrm{ms}$, as shown in Figs.~\ref{fig:2}(a4, b4).

In contrast, when an LG-G pulse sequence is performed under the STIRAP process, the relatively small spatial overlap between the pump LG beam and $\Psi_1$ requires a higher peak amplitude, $\nu_{\max}=500\,\hbar\omega$, in order to achieve $100\%$ population transfer from $\ket{1}$ to $\ket{2}$. In this case, the annular intensity profile of the LG pump beam generates an effective repulsive potential at the periphery of $\Psi_1$ [see Eq.~\eqref{eq:2}], causing the transfer to initiate predominantly from the outer region of $\Psi_1$ to $\Psi_2$ as shown in Figs.~\ref{fig:2}(d2, d3). Similar to the previous case, the spatially dependent Raman coupling is weak near the center of the trap and stronger at the periphery, resulting in a population enrichment observed in the peripheral region of $\Psi_2$ [Fig. \ref{fig:2}(e2)]. This population density combines with the phase factor $e^{-\mathrm{i}(l_1-l_2)\varphi}$ in the coupling term of Eq. \eqref{eq:1}, leading to the nucleation of a vortex [see Figs. \ref{fig:2}(e3, e4)]. The transfer of angular momentum to $\Psi_2$ during the dynamics, as well as the evolution of the number of particles in each hyperfine state, is illustrated in Fig.~\ref{fig:2}(f). Furthermore, similar to the G-LG sequence, the nonuniform, spatially dependent particle transfer, together with slight density deformation induced by the in-plane magnetic field, excites scissors and quadrupole modes in $\Psi_2$ (see Appendix~\ref{app:modes}) \cite{ferrier-barbut_scissors_2018, van_bijnen_collective_2010, pang_bose-einstein_2025, he_collective_2025, qu_scissors_2023, kasamatsu_quadrupole-scissors_2004}. \par

We further find that STIRAP can achieve efficient angular momentum transfer in the SF phase regime, irrespective of the orientation of the external magnetic field relative to the beam propagation direction (not shown here). However, the parallel alignment minimizes the excitation of scissors and quadrupole modes, thereby facilitating angular momentum transfer with reduced fluctuations.

\subsection{Droplet phase}
\label{subsec:Droplet}

\begin{figure*}[t] 
    \includegraphics[width=\textwidth]{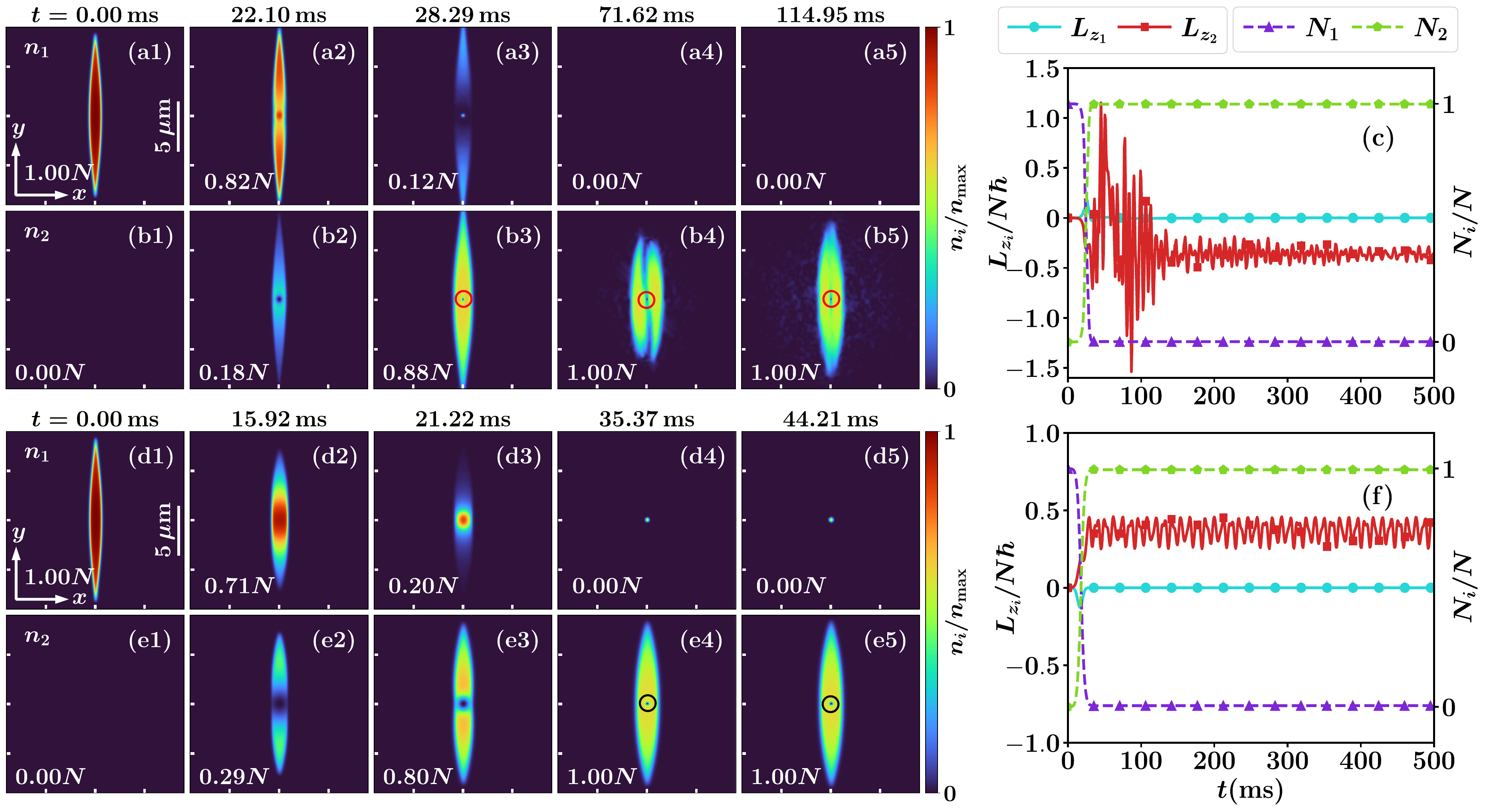}
    \caption{Snapshots of the condensate density distributions during the STIRAP-induced population transfer dynamics in the droplet phase with dipole polarization along the $y$-axis. Panels (a1–a5) and (b1-b5) represent the density profiles of states $\ket{1}$ and $\ket{2}$, respectively, for the G-LG pulse sequence, while panels (d1–d5) and (e1-e5) show the density profiles of states $\ket{1}$ and $\ket{2}$, respectively, for the LG-G pulse sequence. The red circle \protect\marker{red}{} and the black circle \protect\marker{black}{} mark the position of the anti-vortex and vortex, respectively. Variation of the angular momentum $L_{z_i}/N\hbar$ and fraction of atom population $N_i/N$ in both states is shown in panel (c) for the G-LG pulse sequence and in panel (f) for the LG-G pulse sequence. 
    }
    \label{fig:3}
\end{figure*}

So far, our study has focused on the transfer of particles and angular momentum within the weak DDI regime. In the strongly interacting DDI regime, dipolar BECs are known to support exotic self-bound and crystalline phases, including quantum droplets. To realize the single-droplet (SD) phase in the lowest hyperfine state $\ket{1}$, we consider a significantly smaller intra-species scattering length, $a_{11}=70a_0$, relative to the corresponding dipolar length $a_{11}^{\mathrm{dd}}=129.28a_0$. For these parameters, and in the absence of pump and Stokes laser fields, the dipolar condensate forms a self-bound droplet characterized by a negative chemical potential ($\mu<0$) and an elongation along the $y$-axis, i.e., the direction of external polarization [see Figs.~\ref{fig:3}(a1) and \ref{fig:3}(d1)]. To ensure that the system remains in the droplet phase after particle transfer from state $\ket{1}$ to $\ket{2}$, we consider $a_{11}=a_{22}=70a_0$. We investigate the impact of STIRAP in the presence of short-range interspecies interactions characterized by $a_{12}=a_{21}=73 a_0$.

In the case of the G-LG pulse sequence with $\nu_{\max}=200\hbar\omega$, the transfer of particles starts from the center of the droplet in the state $\ket{1}$ [see Fig. \ref{fig:3}(a2)], as a consequence of the effective repulsive potential created by the Gaussian beam [see Eq. \eqref{eq:2}]. The vanishing Raman coupling strength at the center, together with the associated phase winding, produces a density depletion and nucleates an anti-vortex in $\Psi_2$ [Fig. \ref{fig:3}(b2, b3)]. Due to the dominant DDI, both components exhibit highly anisotropic density distributions that are elongated along the direction of the external magnetic field (SD) [Figs. \ref{fig:3}(a3, b3)]. Furthermore, the strong repulsive interaction between species, satisfying $a_{12}>\sqrt{a_{11}a_{22}}$, generates an additional repulsive potential within the droplet in the state $\ket{2}$, causing it to split into two fragments around $30~\mathrm{ms}$ [Fig. \ref{fig:3}(b4)]. Following complete population transfer from $\ket{1}$ to $\ket{2}$, the system gradually relaxes, and the two fragments eventually merge into a single self-bound droplet at approximately $71.62~\mathrm{ms}$ [Fig. \ref{fig:3}(b5)]. \par

In Fig. \ref{fig:3}(c), we have shown the time evolution of the number of particles and the corresponding angular momentum of the condensates. During the dynamics of population transfer, the fragmentation and subsequent merging of the droplet in the state $\ket{2}$ at later times give rise to significant fluctuations in $L_{z_2}$. However, as the system relaxes to an SD after the completion of particle transfer, the oscillation in $L_{z_2}$ decreases and the system carries an average angular momentum of approximately $-0.4 N\hbar$. The small residual fluctuations in $L_{z_2}$ at these later times arise from the anisotropic density distribution of the SD state in the hyperfine level $\ket{2}$, which breaks the rotational symmetry and consequently excites scissors and quadrupole modes (see the Appendix \ref{app:modes}).\par

For the case of the LG-G pulse sequence, we consider $\nu_{\max}=800\hbar \omega$ for the complete transfer of atoms from state $\ket{1}$ to $\ket{2}$, similar to the SF case. In this case, the particle transfer is initiated away from the droplet center along its axial direction due to the overlap between the annular intensity profile of the LG beam and the elongated droplet [see Fig. \ref{fig:3}(d2)]. Consequently, the axial extent of the droplet in state $\ket{1}$ decreases during the transfer process [see Fig. \ref{fig:3}(d3)], leading to a reduction in its anisotropy and the emergence of a localized, nearly symmetric density peak. As the transfer proceeds, the combined effects of the weak Raman coupling at the center, the phase winding imparted by the light fields, and the repulsive interspecies contact interaction induce a central density depletion and facilitate vortex nucleation within the droplet in the hyperfine state $\ket{2}$ [Fig. \ref{fig:3}(e2-e5)]. Unlike the case of G-LG, where the highly anisotropic density distribution of atoms in state $\ket{1}$ generates an anisotropic repulsive barrier that fragments the droplet in state $\ket{2}$, the nearly symmetric density distribution at the center of the condensate in state $\ket{1}$ produces an approximately isotropic repulsive potential for atoms in state $\ket{2}$, facilitating the stable formation of the vortex core \cite{richaud_vortices_2020} without droplet fragmentation. Consequently, $L_{z_2}$ gradually increases and does not exhibit significant fluctuations in the early stages of the dynamics, as shown in Fig. \ref{fig:3}(f). As the particles transferred from $\ket{1}$ to $\ket{2}$, the system acquires an average angular momentum of $L_{z_2}=0.4N\hbar$. The small fluctuations in $L_{z_2}$ observed at later times arise from the excitation of scissors and quadrupole modes, induced by the broken rotational symmetry associated with the anisotropic density profile of the SD in the hyperfine state $\ket{2}$ (see Appendix \ref{app:modes} for further details).

So far, we have focused on a SD configuration in which the magnetic field is aligned perpendicular ($y$-axis) to the direction of beam propagation ($z$-axis). We further find that the droplet becomes dynamically unstable when the magnetic field is aligned parallel to the propagation direction. In this parallel configuration, the vortex core tends to deplete a density tube dominated by attractive head-to-tail dipolar interactions, leading to droplet fragmentation or vortex-line instability (not shown here). In contrast, when the magnetic field is aligned perpendicular to the beam propagation direction, the vortex core predominantly removes a region within the droplet characterized by repulsive dipolar interactions, thereby enhancing the stability of the vortex state. Similar findings were reported in Ref. \cite{PhysRevA.98.023618} for vortex lines imprinted parallel and perpendicular to the direction of the external magnetic field.

\subsection{Supersolid phase}
\label{subsec:supersolid}
\begin{figure*}[t]
    \includegraphics[width=\textwidth]{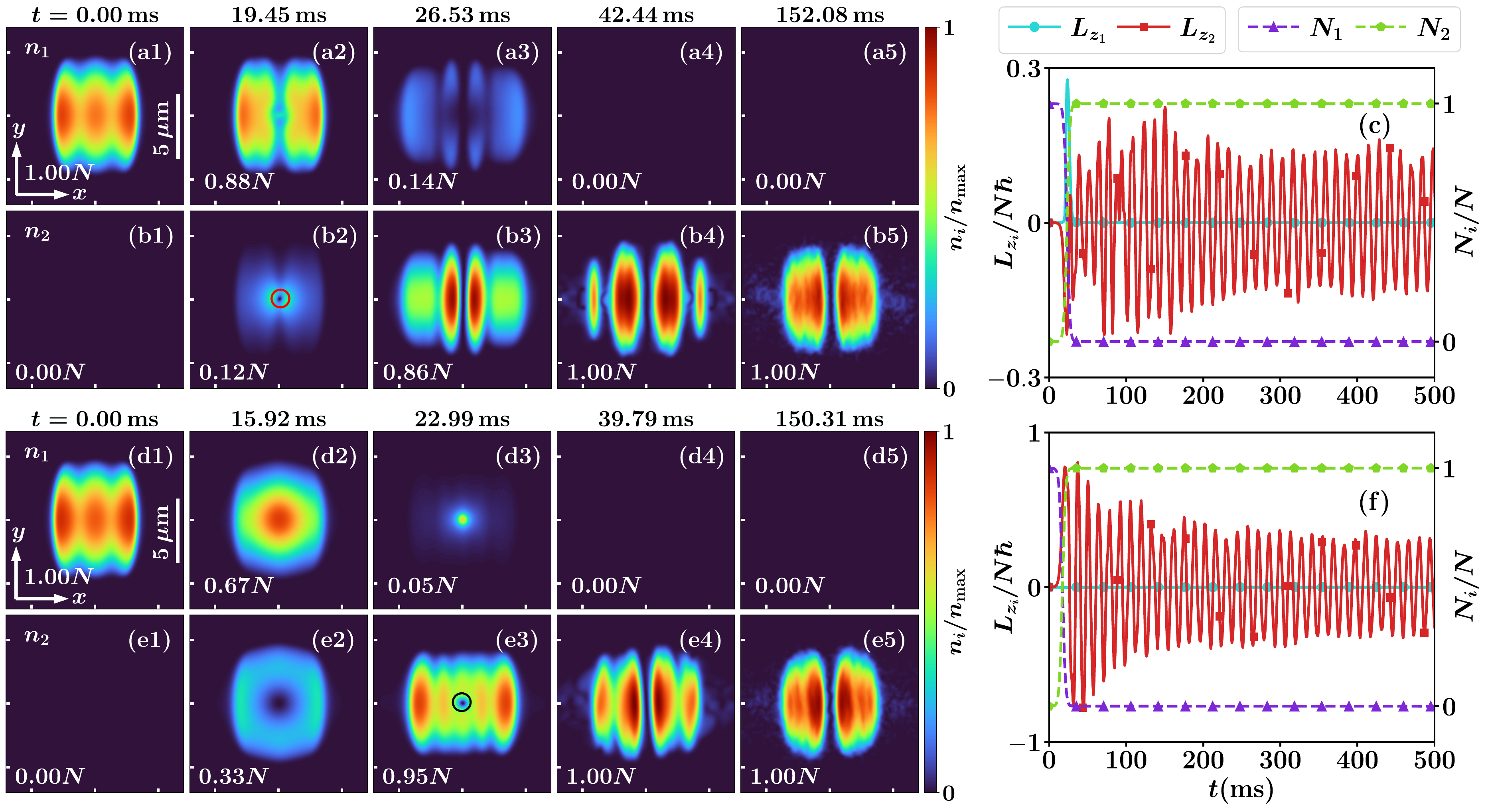}
    \caption{Snapshots of the condensate density distributions following the STIRAP-induced population transfer dynamics in the supersolid phase with dipole polarization along the $y$-axis. Panels (a1–a5) and (b1-b5) represent the density profiles of states $\ket{1}$ and $\ket{2}$, respectively, for the G-LG pulse sequence, while panels (d1–d5) and (e1-e5) show the density profiles of states $\ket{1}$ and $\ket{2}$, respectively, for the LG-G pulse sequence. The red circle \protect\marker{red}{} in (b2) and the black circle \protect\marker{black}{} in (e3) mark the position of the anti-vortex and vortex, respectively. Variation of the angular momentum $L_{z_i}/N\hbar$ and fraction of atom population $N_i/N$ in both states is shown in panel (c) for the G-LG pulse sequence and in panel (f) for the LG-G pulse sequence. 
    }
    \label{fig:4}
\end{figure*}

In the supersolid regime, realized for $\kappa = 2.5$ with intra-species scattering lengths $a_{11}=84 a_0$ and $a_{22}=72 a_0$ and inter-species scattering length $a_{12}=a_{21}=82 a_0$, the ground-state solution of Eq.~\eqref{eq:1} corresponds to a supersolid configuration in state $\ket{1}$.

 For the G-LG pulse sequence with $\nu_{\max}=100\hbar \omega$, the effective repulsive potential in the central region due to the Gaussian pump [see Eq. (\ref{eq:2})] initiates transfer of atoms from the central region of $\ket{1}$. As a result, we observe the density depletion at the center in $\ket{1}$, leading to the formation of four droplets, as shown in Figs.~\ref{fig:4}(a1-a3). Rather than preserving the original droplet symmetry, the supersolid in $\ket{1}$ becomes anisotropic during the population transfer. The density-modulated structure in $\Psi_2$ gradually develops together with an anti-vortex around $t\approx19.45$ ms, [see Fig.~\ref{fig:4}(b2)]. The spatially inhomogeneous particle transfer, together with density deformation induced by the in-plane magnetic field, excites the collective modes in $\Psi_2$. These excited collective modes make the supersolid elongated along the $x$-axis, resulting in an emergence of a weaker connection of low density along the $y$-direction between the droplets on both sides around $\approx26$ ms, as shown in Fig.~\ref{fig:4}(b3). The low-density weak-link region corresponds to a reduced superfluid density between the droplets, which does not provide sufficient phase coherence to sustain the previously nucleated anti-vortex, and eventually the anti-vortex decays.

Similarly to the SF and droplet cases, atom-transfer begins at the periphery of $\Psi_1$ for the LG-G beam sequence with $\nu_{\max}=500\hbar \omega$, and the density modulation in $\Psi_1$ vanishes gradually. As a result, the supersolid structure in $\Psi_1$ temporarily collapses to a more uniform density profile before re-establishing the modulation, as seen in Figs.~\ref{fig:4}(e2, e3). The intermediate configurations reveal a journey of superfluid-to-supersolid transitions rather than a simple droplet. In $\Psi_2$, the density modulation develops from the edge toward the center. The presence of $\Psi_1$ at the center, together with the effective repulsive potential due to the Gaussian beam in the central region [see Eq. (\ref{eq:2})], makes $\Psi_2$ favorable to host a vortex at the center at around $t\approx22.99$ ms [see Fig.~\ref{fig:4}(e3)]. Similar to the G-LG pulse sequence, the excited collective modes significantly reduced the background superfluid density between the droplets in $\Psi_2$. Consequently, the superfluid component becomes too weak to support a stable quantized circulation, and the vortex does not survive, as shown in Fig.~\ref{fig:4}(e4).

The early-time dynamics of the condensate for both pulse sequences exhibit a reduced population in the central region of $\Psi_{2}$. Over longer times, the density of the supersolid state, $\Psi_2$, becomes a disconnected, two-lobed density-modulated structure, as shown in Figs.~\ref{fig:4}(b5, e5). Although the long-time configuration is qualitatively similar in both pulse sequences, the transient pathways differ substantially; elongated droplet formation dominates in the G-LG, whereas transient suppression of density modulation precedes reorganization in the LG-G case. As expected, the angular momentum, $|L_z|$, initially approaches a unit value over a few milliseconds during which a vortex is formed. Then the angular momentum continues to oscillate around zero for a long time, indicating repeated $\pi$-phase jumps and loss of vortex from the condensate due to the significant reduction of background superfluid density between the droplets in $\Psi_2$, as shown in Figs.~\ref{fig:4}(c, f). The oscillation of angular momentum and the resulting anti-vortex and vortex dynamics are accompanied by the excitation of the scissors and quadrupole modes of the condensate, discussed in the Appendix~\ref{app:modes}.

\begin{figure}[t]
    \centering
    \includegraphics[width=\linewidth]{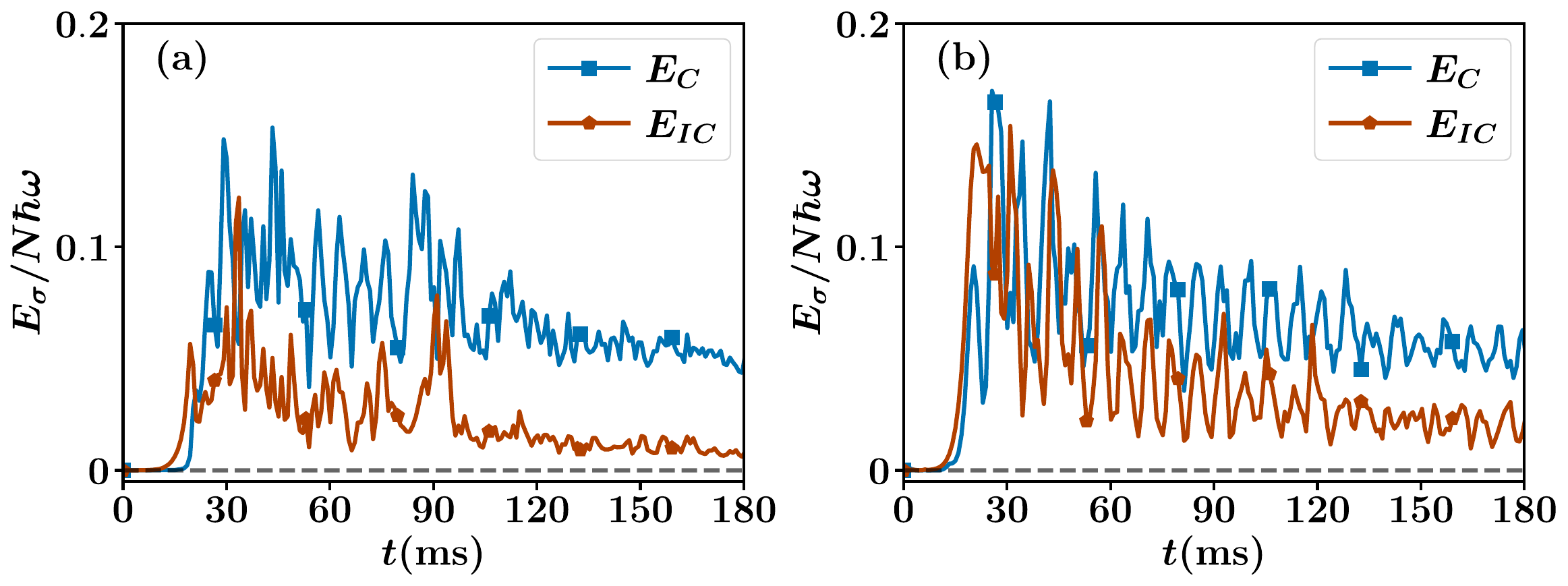} 
    \caption{Time evolution of the compressible kinetic energy $E_C$ and incompressible kinetic energy $E_{IC}$ of state $\ket{2}$ in the supersolid phase for (a) the G-LG pulse sequence and (b) LG-G pulse sequence.
    }
    \label{fig:5}
\end{figure}

We can understand the development of density modulation and the nucleation and decay of vortex (anti-vortex) in the supersolid structure of condensate from the rotational ($E_{IC}$) and irrotational ($E_{C}$) parts of the kinetic energy \cite{escartin_merging_2022, overeng_topological_2025}. For this purpose, we use the density-weighted velocity field $\mathbf{u}= \frac{\hbar}{m}\sqrt{\rho} \nabla \phi,$ where $\rho$ and $\phi$ denote the condensate density and phase, respectively. The kinetic energy contributions are given by
\begin{align*}
E_C &= \frac{m}{2}\int d\mathbf{r}\, |\mathbf{u}^{C}|^{2}, \\
E_{IC} &= \frac{m}{2}\int d\mathbf{r}\, |\mathbf{u}^{IC}|^{2},
\end{align*}
where the compressible velocity field $\mathbf{u}^{C}$ is curl-free, while the incompressible component $\mathbf{u}^{IC}$ is divergence-free.

The compressible and incompressible parts of the kinetic energy are associated with sound waves (density fluctuations) and vortices, respectively. The compressible $E_C$ and incompressible $E_{IC}$ energies rise rapidly at the initial stage, as shown in Figs.~\ref{fig:5}(a) and \ref{fig:5}(b), for both pulse sequences of G-LG and LG-G. During this stage, density modulations develop, and phase gradients build up, marking the onset of the supersolid state. The peak in $E_C$ reveals a significant rearrangement of density as the modulated structure forms, while the increase in $E_{IC}$ suggests the temporary development of a vortex or an anti-vortex in the region of low superfluid fraction around $\approx89.3$ ms to $\approx93.72$ ms for the G-LG case. In the case of the LG-G beam sequence, the two side lobes of the supersolid come close and go apart. This process forms a transient vortex or an anti-vortex, leading to a large oscillation of $E_{IC}$ for $\sim 60$ ms.

Later, at $t \gtrsim 100~\mathrm{ms}$), $E_{IC}$ becomes close to zero, indicating loss of vortex or anti-vortex. In contrast, the oscillation of $E_C$ decays slowly, indicating persistent density modulation. It also suggests that part of the vortex (anti-vortex) energy is redistributed into phonon-like modes. This implies that the supersolid structure remains intact with its two lobes but without a sustained vortex or an anti-vortex. In a nutshell, incompressible energy profiles indicate that the vortex (or anti-vortex), which formed at an early stage of population transfer through light, eventually disappears in supersolid, unlike in superfluid or droplet cases, as discussed earlier.

In contrast to the quasi-2D supersolid with magnetic-field polarization along the $y$-axis, where the vortex (or anti-vortex) leaves the condensate, the supersolid with $z$-axis polarization exhibits long-lived anti-vortex states. In this case, the anti-vortex remains trapped within the condensate, while the transferred component simultaneously develops a well-defined density-modulated supersolid structure. The enhanced stability arises from the alignment of both the magnetic field polarization and the LG-beam propagation along the $z$-direction.

\subsubsection*{Effect of parallel alignment of the magnetic field and the beam propagation direction}
\label{subsec:SupZdir}

\begin{figure*}[t] 
    \includegraphics[width=\textwidth]{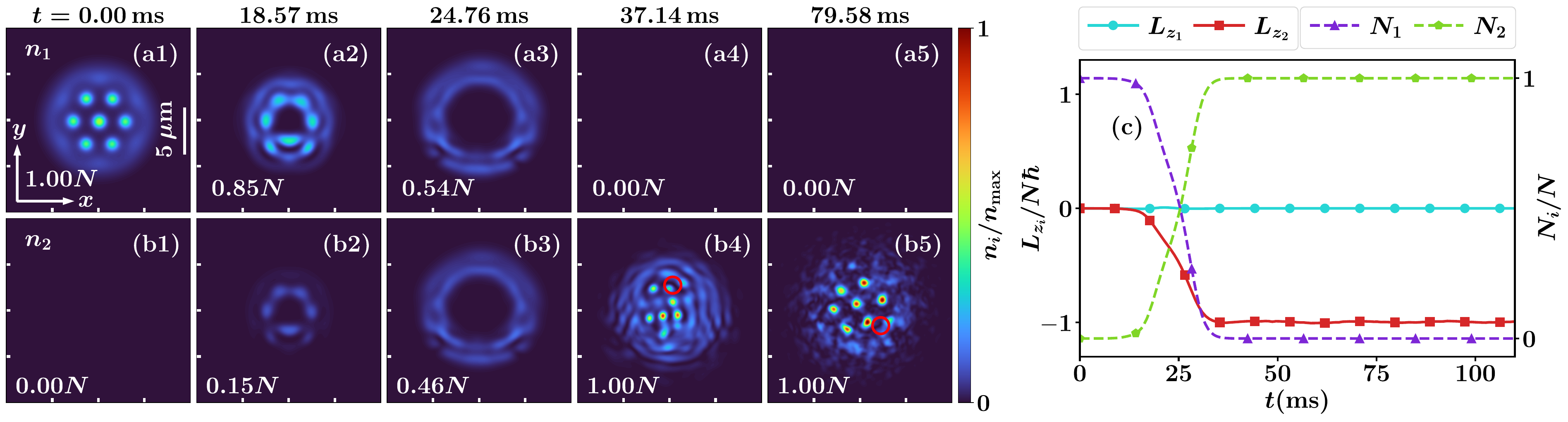}
    \caption{Snapshots of the condensate density distributions during the STIRAP-induced population transfer dynamics in the supersolid phase with dipole polarization along the $z$-axis under G-LG pulse sequence. Panels (a1–a5) and (b1-b5) represent the density profiles of states $\ket{1}$ and $\ket{2}$, respectively.  The red circle \protect\marker{red}{} in (b4) and (b5) marks the position of the anti-vortex. Variation of the angular momentum $L_{z_i}/N\hbar$ and fraction of atom population $N_i/N$ in both states is shown in panel (c) for the G-LG pulse sequence.
    }
    \label{fig:6}
\end{figure*}

When the external magnetic field is aligned parallel to the beam propagation direction, i.e., along the $z$-axis, the trap aspect ratio $\lambda$ must be reduced to enhance the attractive DDI and realize the initial supersolid state. In this configuration, the supersolid state in state $\ket{1}$ is realized for $\lambda = 2.95$, $\kappa = 1$, and $N = 10^{5}$, with intra-species scattering lengths $a_{11}=91a_{0}$ and $a_{22}=71a_{0}$, and inter-species scattering lengths $a_{12}=a_{21}=90a_{0}$. Due to the relatively small value of $\lambda$, the quasi-2D approximation is no longer valid in this regime. Consequently, we solve Eq.~\eqref{fieldequation2} using the full three-dimensional numerical approach within the beyond-mean-field framework.

For the G-LG pulse sequence with $\nu_{\max}=200\hbar\omega$, atoms are transferred from the central region of the supersolid in state $\ket{1}$ to state $\ket{2}$. As the transfer starts, density-modulated droplets still remain in $\ket{1}$ while gradually emerging in $\ket{2}$ at around $\approx18.57$ ms [see Figs.~\ref{fig:6}(a2) and \ref{fig:6}(b2)]. At intermediate times [see Figs.~\ref{fig:6}(a3) and \ref{fig:6}(b3)], both states $\ket{1}$ and $\ket{2}$ contain modulated superfluid background at around $\approx24.76$ ms. With increasing population transfer, this background gradually evolves into a well-defined supersolid structure [see Fig.~\ref{fig:6}(b4)] at around $\approx37.14$ ms. At the same time, an anti-vortex is created in the low-density superfluid background of $\ket{2}$ through the orbital angular momentum carried by the LG beam [see Figs.~\ref{fig:6}(b4) and \ref{fig:6}(b5)]. 

During transfer [see Figs.~\ref{fig:6}(b2), \ref{fig:6}(b3), and \ref{fig:6}(b5)], the size of the supersolid in state $\ket{2}$ grows from the center to the outer region and finally becomes a complete supersolid. The anti-vortex remains trapped inside the supersolid, as the light-matter coupling weakens. The angular momentum remains nearly constant for more than $100$ ms, as shown in Fig.~\ref{fig:6}(c). Since the angular momentum $L_z$ commutes with the Hamiltonian because the magnetic field and the LG beam propagation direction are the same, the transferred angular momentum remains conserved throughout the evolution. As a result, no noticeable collective excitation is observed during the evolution.

Similar behavior is observed for the LG-G pulse sequence, where the transferred vortex remains trapped within the supersolid, and the angular momentum is sustained over long times.

\section{Conclusion}
\label{sec:conclusion}

In this paper, we have demonstrated that the orbital angular momentum of an LG beam can be coherently transferred to a dipolar BEC via STIRAP, but the extent to which it survives and the associated vortex nucleation dynamics it engenders are profoundly governed by the underlying interaction-driven dipolar condensate's phases and by the alignment of the dipole polarization with respect to the direction of propagation of the LG beam. Our results demonstrate that the combination of long-range anisotropic DDI, trap anisotropy, and coherent light-matter coupling enables rich and phase-sensitive vortex nucleation dynamics and collective excitation phenomena.\par

In the superfluid phase, angular momentum is efficiently transferred to the condensate irrespective of the orientation of the dipole polarization relative to the beam propagation direction. This leads to the nucleation of a long-lived quantized vortex. In contrast, for the DDI-dominated phases, namely the droplet and supersolid phases, the orientation of dipole polarization significantly affects the transfer of angular momentum. In the droplet phase, when the magnetic field is aligned perpendicular to the direction of the beam propagation, the vortex remains pinned within the density profile but exhibits complex dynamics, accompanied by the fragmentation and subsequent recombination of droplets. Consequently, the droplet retains only a fraction of the imparted angular momentum, while small fluctuations emerge due to the excitation of scissors and quadrupole modes. However, for the parallel configuration of the magnetic field and beam propagation direction, the vortex becomes dynamically unstable and induces droplet splitting. In the supersolid phase, the spontaneous breaking of continuous translational symmetry introduces density modulation and reduces inter-droplet phase coherence, yielding a qualitatively distinct evolution of the vortex compared to both superfluid and droplet phases. Unlike in the droplet, where the vortex remains confined in the perpendicular configuration, the modulated density structure and reduced superfluid fraction of the supersolid phase facilitate vortex delocalization and eventually escape from the condensate, resulting in a vanishing average angular momentum. However, reorienting the magnetic polarization along the beam propagation direction restores angular momentum conservation and stabilizes the vortex within the supersolid, highlighting the crucial role of geometric alignment between the dipole polarization and the LG beam propagation direction in coherent light engineering.

Further, we examined the role of collective excitations, such as scissors and quadrupole modes, in the population-transferred anisotropic dipolar condensates. These modes are excited by the combined effect of dipolar anisotropy and trap deformation, reflecting the broken rotational symmetry of the system. Their behavior evolves systematically from robust long-lived oscillations in the superfluid to rapid suppression in the supersolid, mirroring the structural evolution of the condensate across the three phases.

Overall, our results establish a versatile and possible route to engineer vortex states and probe collective excitations in dipolar Bose gases via coherent light-matter coupling, where the underlying dipolar phases as well as the relative orientation between the dipole polarization and the LG beam propagation direction play a crucial role in determining the outcome. This work not only enhances our understanding of vortex nucleation, angular momentum retention, and collective dynamics in dipolar gases but also opens new avenues for coherent-state engineering in strongly correlated systems. The theoretical framework developed here is readily extendable to spin-orbit-coupled Bosonic gases \cite{PhysRevA.110.033325, Saboo_2025} and other strongly correlated dipolar systems \cite{bigagli_observation_2024, langen_dipolar_2025, matteo_self-bound_2025}, offering promising directions for ongoing and future experiments in ultracold quantum matter.

\section{Acknowledgements} 
We acknowledge the National Supercomputing Mission (NSM) for providing computing resources of PARAM Shakti at IIT Kharagpur, which is implemented by C-DAC and supported by the Ministry of Electronics and Information Technology (MeitY) and Department of Science and Technology (DST), Government of India. D.S.~acknowledges the MHRD Government of India for the research fellowship. H.S.G.~and A.S.~gratefully acknowledge the support from the Prime Minister's Research Fellowship (PMRF), India. S.M.~gratefully acknowledges the financial support from the Science and Engineering Research Board (SERB) MATRICS project under Grant No.~MTR/2023/000457.

\appendix
\section{Equation of Motions}
\label{app:eom}
In this section, we provide a detailed derivation of the Raman-coupled extended Gross-Pitaevskii Eq.~\eqref{eq:1} mentioned in the main text, Sec.~\ref {sec:theory}. Let $\hat{\Psi}_i^\dagger$ and $\hat{\Psi}_i$ be the creation and annihilation operators, respectively, for atoms in the state $\vert i \rangle$. Let us consider that the Hamiltonian $\hat{H}$ for interacting bosons confined in a trapping potential, written in a frame rotating at the frequency of the applied laser fields within the rotating-wave approximation, is given by \cite{mukherjee_dynamics_2021, ghosh_dastidar_pattern_2022}:

\begin{align*}
\hat{H} &= \hat{H}_{0} + \hat{H}_{\mathrm{c}} + \hat{H}_{\mathrm{dd}} + \hat{H}_{\mathrm{LM}} \\&  + \int d\mathbf{r} \hbar \delta \hat{\Psi}_3^{\dagger}         (\mathbf{r},t) \hat{\Psi}_{3}(\mathbf{r}, t).
\end{align*}

Where $\hat{H}_{0} = \int d\mathbf{r}\sum_{i=1}^{2} \hat{\Psi}_{i}^{\dagger}(\mathbf{r}, t) \hat{h}_{i} \hat{\Psi}_{i}(\mathbf{r}, t)$ with $\hat{h}_{i}=-\hbar^2\nabla^2/2m_i +V_i(\mathbf{r})$. $\hat{H}_{\mathrm{c}}$ and $\hat{H}_{\mathrm{dd}}$ are intra- and inter contact and dipole-dipole interaction terms, respectively.

$$\hat{H}_{\mathrm{c}} = \dfrac{1}{2} \int d\mathbf{r} \sum_{i,j=1}^{2} U_{ij} \hat{\Psi}_{i}^{\dagger}(\mathbf{r}, t) \hat{\Psi}_{j}^{\dagger}(\mathbf{r}, t) \hat{\Psi}_{j}(\mathbf{r}, t) \hat{\Psi}_{i}(\mathbf{r}, t)$$

\begin{align*}
\hat{H}_{\mathrm{dd}} =& \dfrac{1}{2} \int d\mathbf{r} \sum_{i,j=1}^{2} \int d\mathbf{r}' \hat{\Psi}_{i}^{\dagger}(\mathbf{r}, t)\hat{\Psi}_{j}^{\dagger}(\mathbf{r'}, t)V_{\mathrm{dd}}^{ij}(\mathbf{r} - \mathbf{r'}) \\
&  \hat{\Psi}_{j}(\mathbf{r'},t)\hat{\Psi}_{i}(\mathbf{r}, t)
\end{align*}

The short-range contact intra- and inter-component interaction strengths are denoted by $U_{ii} = 4 \pi \hbar^{2} a_{ii} / m_{ii}$ and $U_{ij} = 2 \pi \hbar^{2} a_{ij} / m_{ij},$ respectively. Here $a_{ii}$ and $a_{ij}$ are intra- and inter-species s-wave scattering lengths, and $m_{ij}= m_{i}m_{j}/(m_{i}+m_{j})$ is the reduced mass. And $V_{\mathrm{dd}}^{ij}(\mathbf{r}-\mathbf{r'}) = C_{\mathrm{dd}}^{ij}(1- 3 \cos^{2}\theta)/4 \pi |\mathbf{r}-\mathbf{r'}|^{3}$ and $C^{ij}_{\mathrm{dd}}=\mu_{0}\mu_{i}^m\mu_{j}^m$, where $\mu_0$ is the vacuum permeability and $\mu_i^m(i=1,2)$ is the magnetic moment, and $\theta$ is the angle between the axis linking the two particles and the dipole polarization direction. $\hat{H}_{\mathrm{LM}}$ designates the coupled LM interaction energy between states $\ket{1}$ and $|3\rangle$, and $|3\rangle$ and $\ket{2}$ with the LG laser modes having $l_{1}$ and $l_{2}$ units of charge, respectively. $\hat{H}_{\mathrm{LM}}$ is given as
\begin{align*}
 \hat{H}_{\mathrm{LM}}=\int d\mathbf{r}  \biggl( \sum_{i=1}^{2} \hbar \Omega_{i}(\mathbf{r}, t) \hat{\Psi}_{3}^{\dagger}(\mathbf{r}, t) \hat{\Psi}_{i}             (\mathbf{r}, t) + \text{h.c.} \biggr)
 \end{align*}

Here, $\Omega_1(\mathbf{r})$ and $\Omega_2(\mathbf{r})$ are the Rabi frequencies of the transitions $\vert 1 \rangle \to \vert 3 \rangle$ and $\vert 3 \rangle \to \vert 2 \rangle$, given by: $ \Omega_1(\mathbf{r}, t) =\mathbf{E}_1(\mathbf{r}, t) \cdot \mathbf{d}_{13}/\hbar$ and
$\Omega_2(\mathbf{r}, t) = \mathbf{E}_2(\mathbf{r}, t) \cdot \mathbf{d}_{32}/\hbar$ where $\mathbf{d}_{13}$ and $\mathbf{d}_{32}$ are the transition dipole moments. We consider $\mathbf{d}_{13} = \mathbf{d}_{23} = \mathbf{d}$. From Eqs.~\eqref{electricfiled} and \eqref{temporal}, we obtain 
\begin{align*}
 \Omega_i(\mathbf{r}, t) =&  
(E_{\text{max}}d/\hbar) 
r_{\perp}^{\lvert l_i \rvert}\exp{-\mathrm{i}(k_{i}z-l_i\varphi - \omega_{i} t)}\\
&\exp{-r_{\perp}^2/w_i^2}\exp{-(t - \tau_{i})^2/T^2}.
\end{align*}

 The last term determines light interaction in state $|3\rangle$, where $\delta$ is the detuning.

Starting from the bosonic commutation relations
\begin{align*}
[\hat{\Psi}_i(\mathbf{r}, t), \hat{\Psi}_j^\dagger(\mathbf{r}', t)] 
&= \delta(\mathbf{r} - \mathbf{r}')\, \delta_{ij}, \\
[\hat{\Psi}_i(\mathbf{r}, t), \hat{\Psi}_j(\mathbf{r}', t)] 
&= 0, \\
[\hat{\Psi}_i^\dagger(\mathbf{r}, t), \hat{\Psi}_j^\dagger(\mathbf{r}', t)] 
&= 0,
\end{align*}
and employing the Heisenberg equation of motion
\begin{align*}
\mathrm{i}\hbar\partial_t\hat{\Psi}_i(\mathbf{r}, t) = [\hat{\Psi}_i(\mathbf{r}, t), \hat{H}],
\end{align*}
We obtain the following dynamical equations:

\begin{align*}
\mathrm{i}\hbar\partial_t &\hat{\Psi}_{i}(\mathbf{r}, t)
= \bigg[\hat{h}_{i} + \sum_{j=1}^{2} U_{ij}\hat{\Psi}_{j}^{\dagger}(\mathbf{r}, t)\hat{\Psi}_{j}(\mathbf{r}, t) \bigg]\hat{\Psi}_{i}(\mathbf{r}, t)\\
& + \sum_{j=1}^{2} \int d\mathbf{r'} V_{\mathrm{dd}}^{ij}(\mathbf{r} - \mathbf{r'})
\hat{\Psi}_{j}^{\dagger}(\mathbf{r'}, t) \hat{\Psi}_{j}(\mathbf{r'}, t) \hat{\Psi}_{i}(\mathbf{r}, t) \\
&+ \hbar \Omega_{i}^{*}(\mathbf{r}, t) \hat{\Psi}_{3}(\mathbf{r}, t),~ i = 1, 2.
\end{align*}

\begin{align*}
    \mathrm{i} \hbar\partial_t \hat{\Psi}_{3}(\mathbf{r}, t) &= \hbar \delta \hat{\Psi}_{3}(\mathbf{r}, t) + \hbar \Omega_{1}(\mathbf{r}, t)\hat{\Psi}_{1}(\mathbf{r}, t) \notag \\
& +\hbar \Omega_{2}(\mathbf{r}, t) \hat{\Psi}_{2}(\mathbf{r}, t)
\end{align*}

Assuming the adiabatic elimination condition, i.e, the value of detuning $\delta$ is large \cite{mukherjee_dynamics_2021}, so the field operator $\hat{\Psi}_3(\mathbf{r}, t)$, i.e. $\mathrm{i} \hbar\partial_t \hat{\Psi}_3(\mathbf{r}, t) = 0$, we can create a dark state, which will show zero loss of the particles from the intermediate state. The adiabatic elimination condition builds a relation among the filed operators: $\delta\hat{\Psi}_3(\mathbf{r}, t) = -\Omega_1(\mathbf{r}, t) \hat{\Psi}_1(\mathbf{r}, t) - \Omega_2(\mathbf{r}, t) \hat{\Psi}_2(\mathbf{r}, t)$. This relation modifies the dynamical equations of $\hat{\Psi}_1(\mathbf{r}, t)$ and $\hat{\Psi}_2(\mathbf{r}, t)$:

\begin{equation*}
\begin{split}
\mathrm{i}\hbar\partial_t &\hat{\Psi}_{i}(\mathbf{r}, t)
= \bigg [\hat{h}_{i} + \sum_{j=1}^{2} U_{ij}\hat{\Psi}_{j}^{\dagger}(\mathbf{r}, t) \hat{\Psi}_{j}(\mathbf{r}, t) \bigg ]\hat{\Psi}_{i}(\mathbf{r}, t)\\
& + \sum_{j=1}^{2} \int d\mathbf{r'} V_{\mathrm{dd}}^{ij}(\mathbf{r} - \mathbf{r'})
\hat{\Psi}_{j}^{\dagger}(\mathbf{r'}, t) \hat{\Psi}_{j}(\mathbf{r'}, t) \hat{\Psi}_{i}(\mathbf{r}, t) \\
& -\frac{\hbar}{\delta}|\Omega_{i}(\mathbf{r}, t)|^2\hat{\Psi}_{i}(\mathbf{r}, t) \\
&  - \frac{\hbar}{\delta}\Omega_{3-i}(\mathbf{r}, t) \Omega_{i}^*(\mathbf{r}, t) \hat{\Psi}_{3-i}(\mathbf{r}, t),~ i = 1,2.
\end{split}
\label{fieldequation}
\end{equation*}
At zero kelvin temperature and in the low-energy $s$-wave scattering limit, the bosonic field operator $\hat{\Psi}_i(\mathbf{r},t)$ can be approximated by a classical complex-valued wavefunction $\Psi_i(\mathbf{r},t)$  within the mean-field framework, neglecting quantum fluctuations \cite{ghosh_dastidar_pattern_2022}.
\begin{equation}
\begin{split}
\mathrm{i}\hbar\partial_t &\Psi_{i}(\mathbf{r}, t)
= \bigg [\hat{h}_{i} + \sum_{j=1}^{2} U_{ij}\Psi_{j}^{*}(\mathbf{r}, t)\Psi_{j}(\mathbf{r}, t) \bigg ]\Psi_{i}(\mathbf{r}, t)\\
& + \sum_{j=1}^{2} \int d\mathbf{r'} V_{\mathrm{dd}}^{ij}(\mathbf{r} - \mathbf{r'})
\Psi_{j}^{*}(\mathbf{r'}, t) \Psi_{j}(\mathbf{r'}, t) \Psi_{i}(\mathbf{r}, t) \\
& -\frac{\hbar}{\delta}|\Omega_{i}(\mathbf{r}, t)|^2 \Psi_{i}(\mathbf{r}, t)  \\
&  - \frac{\hbar}{\delta}\Omega_{3-i}(\mathbf{r}, t) \Omega_{i}^*(\mathbf{r}, t) \Psi_{3-i}(\mathbf{r}, t),~ i = 1,2.
\end{split}
\label{fieldequation1}
\end{equation}

However, a dipolar Bose gas exhibits interaction-driven density-modulated phases, including droplet and supersolid states. The conventional mean-field description alone is insufficient to stabilize these phases against collapse \cite{lima_beyond_2012, petrov_quantum_2015, boudjemaa_theory_2014} induced by the dipole-dipole interaction. Consequently, the beyond-mean-field Lee-Huang-Yang (LHY) correction becomes essential for stabilizing droplets and supersolid phases. We include the LHY correction term $\Delta \mu_{i}$ in Eq.~\eqref{fieldequation1}, which accounts for the correction in the chemical potential \cite{bisset_quantum_2021, halder_two-dimensional_2023,halder_induced_2024, smith_quantum_2021, zhang_metastable_2024, bland_alternating-domain_2022, scheiermann_catalyzation_2023}.

\begin{equation}
\begin{split}
\mathrm{i}\hbar\partial_t &\Psi_{i}(\mathbf{r}, t)
= \bigg [\hat{h}_{i} + \sum_{j=1}^{2} U_{ij}\Psi_{j}^{*}(\mathbf{r}, t)\Psi_{j}(\mathbf{r}, t) \bigg ]\Psi_{i}(\mathbf{r}, t)\\
& + \sum_{j=1}^{2} \int d\mathbf{r'} V_{\mathrm{dd}}^{ij}(\mathbf{r} - \mathbf{r'})
\Psi_{j}^{*}(\mathbf{r'}, t) \Psi_{j}(\mathbf{r'}, t) \Psi_{i}(\mathbf{r}, t) \\
& -\frac{\hbar}{\delta}|\Omega_{i}(\mathbf{r}, t)|^2 \Psi_{i}(\mathbf{r}, t) + \Delta \mu_{i} \Psi_{i}(\mathbf{r}, t) \\
&  - \frac{\hbar}{\delta}\Omega_{3-i}(\mathbf{r}, t) \Omega_{i}^*(\mathbf{r}, t) \Psi_{3-i}(\mathbf{r}, t),~ i = 1,2.
\end{split}
\label{fieldequation2}
\end{equation}

 Along with this tight confinement of the condensate in the $z=0$ plane and equal frequency ($\omega_1=\omega_2$) of the laser pulses, the Eq.~\eqref{fieldequation2} reduces to Eq.~\eqref{eq:1}.

\begin{figure*}[t]
    \includegraphics[width=\textwidth]{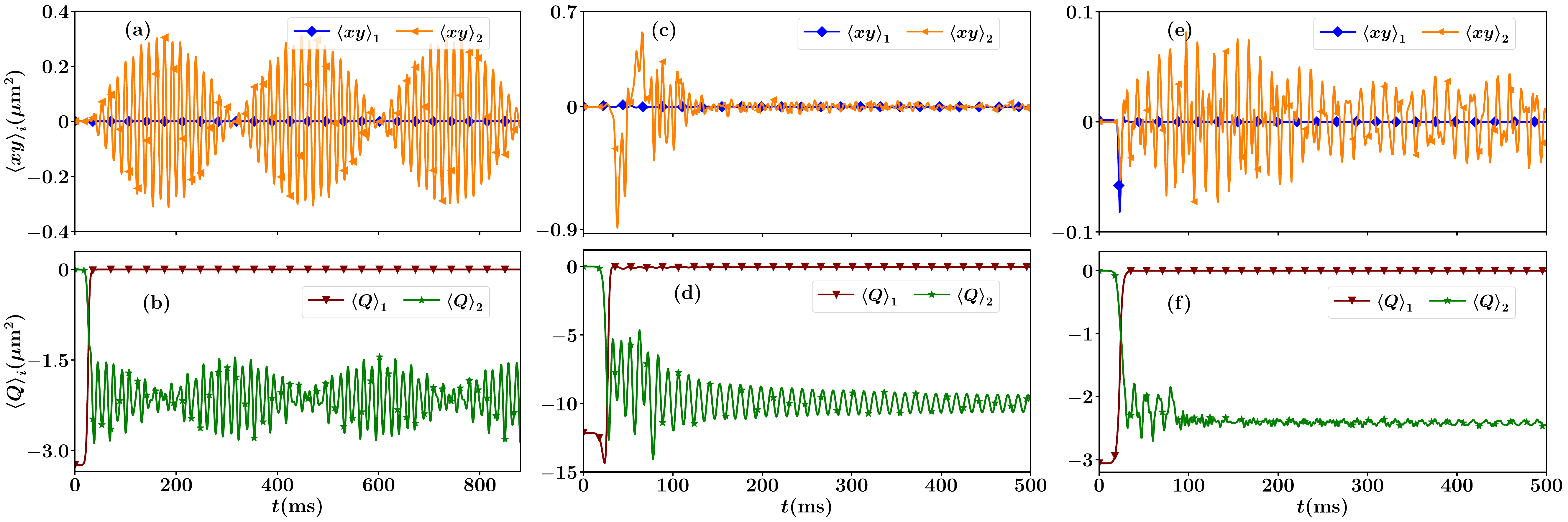}
    \caption{Time evolution of the scissors $\expval{xy}_i$ and quadrupole $\langle Q \rangle_i$ modes under the G-LG pulse sequence for the (a, b) superfluid, (c, d) droplet, and (e, f) supersolid phases, respectively, with the dipole polarization oriented perpendicular to the LG-beam propagation direction. 
    }
    \label{fig:7}
\end{figure*}

\section{Quasi-2D LHY Correction}
\label{app:lhy}

In this section, we provide a detailed derivation of the quasi-2D form of LHY correction Eq.~\eqref{eq:3} mentioned in the main text, Sec.~\ref {sec:theory}. The correction to the chemical potential $\Delta\mu_i$ resulting from the effect of quantum fluctuations for a binary dipolar mixture in 3D is given by \cite{bisset_quantum_2021, smith_quantum_2021, bland_alternating-domain_2022, halder_two-dimensional_2023, scheiermann_catalyzation_2023, zhang_metastable_2024, halder_induced_2024}.  
\begin{align}
&\Delta\mu_i = \frac{m^{3/2}}{3\sqrt{2}\pi^2 \hbar^3} \sum_{\pm} \int_0^{1} du \nonumber \\
&\quad Re\left( \Bigg[\Tilde{V}_{ii}^{int}(u) \pm \frac{(-1)^{i-1}\delta \Tilde{V}_{ii}^{int}(u) + 2\Tilde{V}_{12}^{int}(u)^2 n_{3-i}}{\sqrt{\delta^2 + 4\Tilde{V}_{12}^{int}(u)^2 n_1 n_2}}\Bigg] I^{3/2}_{E \pm}\right)
\label{3dlhy}
\end{align}
with
$\delta = n_1 \Tilde{V}_{11}^{int}(u) - n_2 \Tilde{V}_{22}^{int}(u), \, \Tilde{V}_{ij}^{int}(u)=\Tilde{g}^{c}_{ij} + \Tilde{V}_{\rm{dd}}^{ij}(u),$ being the Fourier transform of the total interaction potential, $I_{E\pm} =  n_1 \Tilde{V}_{11}^{int}(u) +  n_2 \Tilde{V}_{22}^{int}(u)  \pm \sqrt{\delta^2 + 4\Tilde{V}_{12}^{int}(u)^2 n_1 n_2}.$ Under strong confinement along the $z$-axis $(\lambda\gg1)$, the axial dynamics of the system can be assumed to be frozen in the axial ground state $\Phi(z) = (1 / \pi l_{z}^2)^{1/4} \exp(-z^{2}/2l^{2}_{z})$ such that the dynamics of the condensate is effectively restricted in the transverse plane. In this scenario the component wave function can be written as $\Psi_{i}(\mathbf{r}_{\perp},z) = \Psi_i(\mathbf{r}_{\perp})\Phi(z)$ and integrating over the $z$-axis we obtain Eq.~\eqref{eq:3} which represents the effective quasi-2D form of the LHY correction for a binary dipolar mixture.

\section{Excitation of scissors and quadrupole modes}
\label{app:modes}
The excitation of scissors \cite{roccuzzo_rotating_2020, ferrier-barbut_scissors_2018, halder_dynamical_2025, odell_exact_2004, guery-odelin_scissors_1999, SSSSdoi:10.1126/science.aba4309} and quadrupole modes \cite{he_collective_2025, qu_scissors_2023} serves as a hallmark signature of superfluidity and emerges as a consequence of broken rotational symmetry. In dipolar condensates, rotational symmetry can be broken by either an anisotropic trapping potential or the intrinsic anisotropy of the DDI. In the present system, the magnetic field is oriented along the $y$-axis, whereas the LG beam propagates along the $z$-axis. Consequently, angular momentum is not conserved, leading to the excitation of scissors and quadrupole modes.\par

In the superfluid phase, the transferred angular momentum per particle in the hyperfine state $\ket{2}$  exhibits small fluctuations around $-\hbar$, as shown in Figs.~\ref{fig:2}(c) and \ref{fig:2}(f). This behavior arises from breaking the rotational symmetry of the system. Although we have considered an isotropic trapping configuration in the $x$-$y$ plane, the presence of the external magnetic field along the $y$-axis induces an anisotropic DDI. Due to the attractive DDI along the $y$-axis, the condensate develops a slightly elongated density distribution along the polarization direction of the external magnetic field. This anisotropic density distribution breaks the rotational symmetry of the system and consequently gives rise to the excitation of scissors and quadrupole modes, as shown in Fig.~\ref{fig:7}(a) and \ref{fig:7}(b). It is evident that the scissors mode, characterized by $\expval{xy}_i=\int \dd{x}\dd{y}xy \abs{\Psi_i(x,y)}^2$, is coupled to the quadrupole mode $\langle Q \rangle$ associated with the operator $Q_i = \left(\omega_x^2 / \omega_y^2 \right)\expval{x^2}_i -\expval{y^2}_i$.\par
In the droplet phase, the stronger DDI gives rise to a highly anisotropic density distribution in the $x$-$y$ plane, resulting in the excitation of scissors and quadrupole modes  \cite{ferrier-barbut_scissors_2018, young-s_dipole-mode_2023} with significantly larger amplitudes, as shown in Figs.~\ref{fig:7}(c) and \ref{fig:7}(d). In contrast, in the SS phase, both the trapping potential and the DDI contribute to the excitation of these collective modes. Although the DDI is stronger than that in the SF phase, the larger confinement frequency $\omega_y$ along the direction of the external magnetic field reduces the effective anisotropy of the density distribution in the SS phase. Consequently, the amplitudes of the scissors and quadrupole oscillations \cite{van_bijnen_collective_2010, young-s_dipole-mode_2023} remain comparable to those observed in the SF phase [Figs. \ref{fig:7}(e) and \ref{fig:7}(f)]. Furthermore, we observe an exponential damping of these oscillations, reflecting the rigid behavior of the droplet and supersolid phases, similar to that reported in ~\cite{halder_dynamical_2025}.

\bibliographystyle{apsrev4-2}
\bibliography{references.bib}
\end{document}